\documentclass{article}

\usepackage[english]{babel}
\usepackage[utf8]{inputenc}
\usepackage[T1]{fontenc}

\usepackage[a4paper,top=2.5cm,bottom=3cm,left=2.5cm,right=2.5cm,marginparwidth=1.75cm]{geometry}

\usepackage{amsmath}
\usepackage[round]{natbib}
\usepackage{graphicx}
\usepackage[colorlinks=true, allcolors=blue]{hyperref}
\usepackage[detect-weight,retain-explicit-plus]{siunitx} %

\usepackage[yyyymmdd]{datetime}

\title{Ambisonic Encoding of Signals From\\ Spherical Microphone Arrays}
\author{Jens Ahrens}
\date{Technical note\footnote{Find more technical notes at \url{http://www.ta.chalmers.se/education/texts-on-acoustics/}}~~v.~1\\[1ex] 
Chalmers University of Technology\\[1ex] 
\texttt{jens.ahrens@chalmers.se}}

\newcommand{\e}{\mathrm{e}}
\renewcommand{\d}{\mathrm{d}}
\renewcommand{\i}{\mathrm{i}}

\begin{document}
\maketitle

\begin{abstract}
This document illustrates how to process the signals from the microphones of a rigid-sphere higher-order ambisonic microphone array so that they are encoded with N3D normalization and ACN channel order and thereby can be used with the standard ambisonic software tools such as SPARTA and the IEM Plugin Suite. A MATLAB script is provided.
\end{abstract}

\section{Introduction}

In the following, we assume that the forward and inverse Fourier transforms over time are defined as
\begin{equation}\label{eq:cft}
     S(\omega)  = \int\displaylimits_{-\infty}^{\infty} s(t)  \ \e^{-\i \omega t} \mathrm{d}t 
\end{equation}
and
\begin{equation}\label{eq:cift}
     s(t) = \frac{1}{2\pi} \int\displaylimits_{-\infty}^{\infty} S(\omega)  \ \e^{\i \omega t} \mathrm{d}\omega \, ,
\end{equation}
where $s(t)$ is the time-domain representation of our signal under consideration and $S(\omega)$ its frequency-domain representation. Note that the algebraic sign of the exponent of the complex exponential is the aspect of importance. Above definitions are compatible with the definition of the discrete Fourier transform in MATLAB\footnote{Note that~\citep[Eq.~(1.6)]{Williams:book1999}, for example, uses a different sign convention, which has the consequence that $h_n^{(1)}(\cdot)$ denotes outward radiating waves there.}.

The sound pressure~$S^\text{\;\!int}(\beta, \alpha, r, \omega)$ in an interior domain i.e.,~the sound pressure inside a spherical domain centered at the coordinate origin that is free of sound sources and free of reflecting boundaries, can be represented in SHs in different ways~\citep{Gumerov:book2005}. To be consistent with the ambisonics formulation, we represent it as~\citep[Sec.~5.5]{Ahrens:binaural_rendering_in_sh}
\begin{equation}\label{eq:sh_decomp_general}
    S^\text{\;\!int}(\beta, \alpha, r, \omega) = 
    \sum_{n=0}^\infty \sum_{m=-n}^n\!\! 4\pi\i^{n} \, \breve{S}_{n,m}(\omega)\,\,j_n\!\left( \omega \frac{r}{c} \right)\,Y_{n,m}(\beta, \alpha) 
\end{equation}
$\breve{S}_{n,m}(\omega)$ is the frequency-domain representation of the ambisonic signals $\breve{s}_{n,m}(t)$ that we aim at extracting from the microphone signals. $\beta$ and $\alpha$ are the colatitude and azimuth angles of a spherical coordinate system, respectively.

When $S^\text{\;\!int}(\beta, \alpha, r, \omega)$ impinges on a rigid sphere of radius $R$ that is centered at the coordinate origin (i.e., when $S^\text{\;\!int}(\beta, \alpha, r, \omega)$ impinges on the surface of our rigid-sphere microphone array), the sound pressure $S(\beta, \alpha, R, \omega)$ on the surface of the sphere is consequently given by
\begin{equation}\label{eq:sh_decomp_general_scat}
    S^{(R)}(\beta, \alpha, R, \omega) =
    \sum_{n=0}^\infty \sum_{m=-n}^n\!\! \breve{S}_{n,m}(\omega)\,\,b_n\!\left( \omega \frac{R}{c}, R \right)\, Y_{n,m}(\beta, \alpha) \ ,
\end{equation}
with radial term 
\begin{eqnarray}\label{eq:radial_filter_3d_wronskian_1}
    b_n\!\left( \omega \frac{R}{c}, R \right) &=& \ \ \, 4\pi\i^{n} \left[ j_n\left(\omega \frac{R}{c} \right) - \frac{j_n'\left(\omega \frac{R}{c} \right)}{{h_n^{(2)}}'\left(\omega \frac{R}{c} \right)} h_n^{(2)}\left(\omega \frac{R}{c} \right) \right] \\ \label{eq:radial_filter_3d_wronskian_2}
    &=& - 4\pi\i^{n}\, 
    \frac{\i}{\left(\omega\frac{R}{c}\right)^2}  \frac{1}{h_n^{\prime\;\!(2)}\!\left( \omega \frac{R}{c} \right)} \ .
\end{eqnarray}
We exploited the Wronskian for spherical Bessel functions~\citep[Eq.~(4.2.13)]{Gumerov:book2005} to convert~\eqref{eq:radial_filter_3d_wronskian_1} into~\eqref{eq:radial_filter_3d_wronskian_2}.\\

\noindent Pressure microphones on the surface of the spherical baffle sense $S^{(R)}(\beta, \alpha, R, \omega)$, which means that we know all quantities in~\eqref{eq:sh_decomp_general_scat} apart from the sought-after ambisonic signals $\breve{S}_{n,m}(\omega)$. We observe that any square-integrable function $X(\beta, \alpha)$ that is defined on the surface of a sphere -- like our sound pressure $S^{(R)}(\beta, \alpha, R, \omega)$ -- can be expanded into SHs coefficients $\mathring{X}_{n,m} $ as~\citep[Eq.~(6.48)]{Williams:book1999}
\begin{equation}\label{eq:sh_general}
    X(\beta, \alpha) =
    \sum_{n=0}^\infty \sum_{m=-n}^n\!\! \mathring{X}_{n,m} \, Y_{n,m}(\beta, \alpha) \ ,
\end{equation}
and that the coefficients $\mathring{X}_{n,m}(\omega)$ of that expansion can be computed via~\citep[Eq.~(6.49)]{Williams:book1999}
\begin{equation}\label{eq:sh_extraction}
    \mathring{X}_{n,m} = 
    \oint_{\Omega \in S^2} X(\beta, \alpha) \, Y_{n,m}(\beta, \alpha) \ \mathrm{d} \Omega 
\end{equation}
because of the orthogonality of $Y_{n,m}(\beta, \alpha)$ with respect to the mode $(n,m)$. $S^2$ denotes the unit sphere (apologies for using the symbol '$S$' again).\\

\noindent In~\eqref{eq:sh_extraction}, we assume that the spherical harmonics $Y_{n,m}(\beta, \alpha)$ are defined as 
\begin{equation}\label{eq:sh_definition_real}
    Y_{n,m}(\beta, \alpha) = (-1)^m\, \sqrt{\frac{2n{+}1}{4\pi} \frac{(n{-}|m|)!}{(n{+}|m|)!}}\,P_n^{|{m}|}(\cos\beta)
\begin{cases}
 \sqrt{2} \sin{|m| \alpha}, \ \forall m < 0\\
 \ \ \ \  \ \ \ 1 \ \ \ \ \ \ \, , \ \forall m = 0\\
 \sqrt{2} \cos{|m| \alpha}, \ \forall m > 0
\end{cases} \ ,
\end{equation}
which corresponds to the definition that is used in the N3D ambisonic format~\citep{Nachbar:AmbiX2011}\footnote{Associated Legendre functions are used with different normalizations. We assume that $P_{n,m}(\mu)$ is defined via the following Rodriguez formula~\citep[Eq.~(2.1.20)-(2.1.21)]{Gumerov:book2005}:
\begin{equation}\nonumber
    P_{n,m}(\mu) = (-1)^m(1-\mu^2)^{m/2} \frac{\d^m}{\d \mu^m} P_n(\mu),  \ \ \forall n\geq 0, \ m\geq m \ ,
\end{equation}
with
\begin{equation}\nonumber
    P_n(\mu) = \frac{1}{2^n n!} \frac{\d^n}{\d \mu^n} (\mu^2 - 1),  \ \ \forall n\geq 0 \ .
\end{equation}
This also the definition that MATLAB and SciPy use.
}. It is also identical to the definition that is used in Politis's \texttt{getSH.m} MATLAB function when called with the argument \texttt{'real'}~\citep{Politis:SH_transform}.\\

Assuming that we have a continuous layer of pressure microphones on the surface of the baffle and comparing~\eqref{eq:sh_general} and~\eqref{eq:sh_decomp_general_scat} allows for establishing the equivalent of~\eqref{eq:sh_extraction} as
\begin{equation}\label{eq:sh_extraction_f}
    \breve{S}_{n,m}(\omega) = \frac{1}{b_n\!\left( \omega \frac{R}{c}, R \right) }
    \oint_{\Omega \in S^2} S^{(R)}(\beta, \alpha, R, \omega) \, Y_{n,m}(\beta, \alpha) \ \mathrm{d} \Omega \ ,
\end{equation}
which extracts the ambisonic signals $\breve{S}_{n,m}(\omega)$ from the microphone signals $S^{(R)}(\beta, \alpha, R, \omega)$. Note that $\frac{1}{b_n\!\left( \omega \frac{R}{c}, R \right)}$ is ill-conditioned at low frequencies and high orders and needs to regularized~\citep{Moreau:AES2006,Bernschutz:PhD2016,Politis:WASPAA2017}.
In practice, the integral in~\eqref{eq:sh_extraction_t} needs to be discretized because $S^{(R)}(\beta, \alpha, R, \omega)$ is measured only at a finite set of discrete microphone positions. This is the cause for the required order truncation as well as spatial aliasing~\citep{Rafaely:TASPL2005}.\\

\noindent It also conceivable to implement~\eqref{eq:sh_extraction_f} in time domain as
\begin{equation}\label{eq:sh_extraction_t}
    \breve{s}_{n,m}(t) = b_n^{(-1)}\!\left( t \frac{R}{c}, R \right) \ast_t
    \oint_{\Omega \in S^2} s^{(R)}(\beta, \alpha, R, t) \, Y_{n,m}(\beta, \alpha) \ \mathrm{d} \Omega \ ,
\end{equation}
whereby $\ast_t$ denotes convolution with respect to time, and $b_n^{(-1)}\!\left( t \frac{R}{c}, R \right)$ is the time-domain representation of $\frac{1}{b_n\!\left( \omega \frac{R}{c}, R \right)}$.\\ 

\noindent Lastly, we need to arrange the channels of the multichannel signal $\breve{s}_{n,m}(t)$ in a standardized order. The \emph{Ambisonic Channel Number} (ACN) is the most popular arrangement. The channel number is given by $n^2+n+m$. An example implementation of the above procedure is available in~\citep{Ahrens:matlab_encoding}. The resulting signals $\breve{s}_{n,m}(t)$ are compatible with the standard ambisonic software tools such as SPARTA\footnote{\url{https://leomccormack.github.io/sparta-site/}} and the IEM Plugin Suite\footnote{\url{https://plugins.iem.at/}}. Make sure that the normalization is set to N3D in the software tool that you are using.

\section{Rendering}

For completeness, we complement the processing pipeline here and illustrate how binaural rendering of the ambisonic signals $\breve{S}_{n,m}(\omega)$ and $\breve{s}_{n,m}(t)$, respectively, is performed, i.e., how the left and right ear signals $B^\text{L,R}(\omega)$ of a virtual listener in the captured sound field are computed. 

Taking into account the conventions used in the present report, binaural rendering is performed as~\citep[Tab.~1]{Ahrens:binaural_rendering_in_sh}
\begin{equation}\label{eq:binaural_rendering}
    B^\text{L,R}(\omega) =\sum_{n=0}^N \sum_{m=-n}^n\!\breve{S}_{n,m}(\omega)\,\,\mathring{H}_{n,m}^\text{L,R}(\omega) \ .
\end{equation}
$\mathring{H}_{n,m}^\text{L,R}(\omega)$ are the SH coefficients of the left and right far-field head-related transfer function $H^\text{L,R}(\beta, \alpha, \omega)$, respectively, defined as
\begin{equation}
    H^\text{L,R}(\beta, \alpha, \omega) =
    \sum_{n=0}^\infty \sum_{m=-n}^n\!\! \mathring{H}_{n,m}^\text{L,R}(\omega)\,Y_n^m(\beta, \alpha)
\end{equation}
$\beta$ and $\alpha$ are the colatitude and azimuth of the sound incidence direction. The MATLAB scripts provided in~\citep{Ahrens:matlab_encoding} also demonstrate how to implement~\eqref{eq:binaural_rendering}.


\begin{thebibliography}{10}
\providecommand{\natexlab}[1]{#1}
\providecommand{\url}[1]{\texttt{#1}}
\expandafter\ifx\csname urlstyle\endcsname\relax
  \providecommand{\doi}[1]{doi: #1}\else
  \providecommand{\doi}{doi: \begingroup \urlstyle{rm}\Url}\fi

\bibitem[Ahrens(2022{\natexlab{a}})]{Ahrens:binaural_rendering_in_sh}
Jens Ahrens.
\newblock {Binaural Audio Rendering in the Spherical Harmonic Domain: A Summary
  of the Mathematics and Its Pitfalls}.
\newblock Technical note v.~2, Chalmers University of Technology,
  2022{\natexlab{a}}.

\bibitem[Ahrens(2022{\natexlab{b}})]{Ahrens:matlab_encoding}
Jens Ahrens.
\newblock {Ambisonic encoding}.
\newblock Chalmers University of Technology,
  \url{https://github.com/AppliedAcousticsChalmers/ambisonic-encoding},
  2022{\natexlab{b}}.

\bibitem[Bernsch\"{u}tz(2016)]{Bernschutz:PhD2016}
Benjamin Bernsch\"{u}tz.
\newblock Microphone arrays and sound field decomposition for dynamic binaural
  recording.
\newblock PhD thesis, Technische Universit\"{a}t Berlin, 2016.

\bibitem[Gumerov and Duraiswami(2005)]{Gumerov:book2005}
Nail Gumerov and Ramani Duraiswami.
\newblock \emph{Fast Multipole Methods for the Helmholtz Equation in Three
  Dimensions}.
\newblock Elsevier, Amsterdam, 2005.

\bibitem[Moreau et~al.(2006)Moreau, Daniel, and Bertet]{Moreau:AES2006}
S{\'e}bastien Moreau, J{\'e}r{\^o}me Daniel, and St{\'e}phanie Bertet.
\newblock {3D} sound field recording with higher order ambisonics – objective
  measurements and validation of a 4th order spherical microphone.
\newblock In \emph{120th Conv.~of the AES}, May 2006.

\bibitem[Nachbar et~al.(2011)Nachbar, Zotter, Deleflie, and
  Sontacchi]{Nachbar:AmbiX2011}
Christian Nachbar, Franz Zotter, Etienne Deleflie, and Alois Sontacchi.
\newblock {ambiX - A Suggested Ambisonics Format (With Comments)}.
\newblock In \emph{Ambisonics Symposium}, Lexington, KY, June 2011.
\newblock
  \url{https://iem.kug.ac.at/fileadmin/media/iem/projects/2011/ambisonics11_nachbar_zotter_sontacchi_deleflie.pdf}.

\bibitem[Politis(2015)]{Politis:SH_transform}
Archontis Politis.
\newblock Spherical harmonic transform library.
\newblock Github,
  \url{https://github.com/polarch/Spherical-Harmonic-Transform}, 2015.

\bibitem[Politis and Gamper(2017)]{Politis:WASPAA2017}
Archontis Politis and Hannes Gamper.
\newblock Comparing modeled and measurement-based spherical harmonic encoding
  filters for spherical microphone arrays.
\newblock In \emph{WASPAA}, pages 224--228, 2017.

\bibitem[{Rafaely}(2005)]{Rafaely:TASPL2005}
Boaz {Rafaely}.
\newblock Analysis and design of spherical microphone arrays.
\newblock \emph{IEEE Trans.~on Speech and Audio Proc.}, 13\penalty0
  (1):\penalty0 135--143, 2005.

\bibitem[Williams(1999)]{Williams:book1999}
Earl Williams.
\newblock \emph{Fourier Acoustics: Sound Radiation and Nearfield Acoustical
  Holography}.
\newblock Academic Press, New York, 1999.

\end{thebibliography}
\end{document}